\documentclass{eptcs}

\usepackage{graphicx}
\usepackage{amsfonts}
\usepackage{latexsym}
\usepackage{amssymb}

        \providecommand{\event}{Gabriel Ciobanu (Ed.): Membrane Computing and
Biologically Inspired Process Calculi 2009} 
        \providecommand{\volume}{15}   
        \providecommand{\anno}{2009}   
        \providecommand{\firstpage}{1} 
        \providecommand{\eid}{1}       
\usepackage{breakurl}        

\title{Drip and Mate Operations Acting in Test Tube Systems and 
  Tissue-like P systems}
\author{Rudolf Freund
\institute{Faculty of Informatics\\
   Vienna University of Technology\\
   Favoritenstr. 9-11, 1040 Vienna, Austria} 
\email{rudi@emcc.at}
\and Marian Kogler   
\institute{Faculty of Informatics\\
   Vienna University of Technology\\
   Favoritenstr. 9-11, 1040 Vienna, Austria} 
\email{marian@emcc.at} 
}
\def\titlerunning{Drip and Mate Operations Acting in Test Tube Systems and 
  Tissue-like P systems}
\def\authorrunning{Rudolf Freund \& Marian Kogler}

\begin{document}
\maketitle

\begin{abstract}
The operations drip and mate considered in (mem)brane computing resemble 
the operations cut and recombination well known from DNA computing. 
We here consider sets of vesicles with multisets of objects on their outside
membrane interacting by drip and mate in two different setups: in test tube 
systems, the vesicles may pass from one tube to another one provided they
fulfill specific constraints; in tissue-like P systems, the vesicles are 
immediately passed to specified cells after having undergone a drip or mate
operation. In both variants, computational completeness can be obtained, 
yet with different constraints for the drip and mate operations.
\end{abstract}

\section{Introduction}

One of the basic operations used in the field of DNA computing was
introduced by Tom Head in \cite{Head87} more than twenty years ago, when he
formalized the operation of splicing, well-known from biology as an
operation on DNA strands: given two strings of symbols $x$ and $y$, the
splicing operation consists of cutting $x$ and $y$ at certain positions
(determined by the splicing rule) and pasting the resulting prefix of $x$
together with the suffix of $y$ as well as pasting the resulting prefix of $y
$ together with the suffix of $x$, respectively. Formally, if we apply the
splicing rule $(u_{1},u_{2};u_{3},u_{4}),$ then the results of splicing $x$
and $y$ are $z$ and $w$ where $x=x_{1}u_{1}u_{2}x_{2}$, $%
y=y_{1}u_{3}u_{4}y_{2}$, and $z=x_{1}u_{1}u_{4}y_{2}$, $%
w=y_{1}u_{3}u_{2}x_{2}$ with $u_{1},u_{2},u_{3},u_{4},x_{1},x_{2},y_{1},y_{2}
$ being strings over a given alphabet $V$. In the case of real DNA
sequences, the alphabet consists of four letters, i.e., \textit{A, C, G, T},
representing the four bases adenine, cytosine, guanine and thymine; the
cutting is realized by restriction enzymes, and the recombination by ligases.

In \cite{DenGatter89}, the range of Turing machines was encoded using
iterated splicing on multisets (sets with multiplicities associated to their
elements). The splicing operation then mainly was used as a basic tool for
building a generative mechanism, called a \textit{splicing system} or 
\textit{H system}, as formalized by Gheorghe P\u{a}un in the following way:
given a set of strings (axioms) and a set of splicing rules, the generated
language consists of the strings obtained in an iterative way by applying
the rules to the axioms and/or to the strings obtained in preceding splicing
steps. If we add the restriction that only strings over a designated subset
of the alphabet are accepted in the language, we obtain an extended H
system. As already shown in \cite{CulikHarju91} and then in \cite{Pixton2000}
for a class of related systems, in that way we can only obtain regular
languages. Yet when considering multisets of strings as already done in \cite%
{DenGatter89} or by adding control mechanisms as used in the area of formal
language theory (e.g., see \cite{DassowPaun89}) like checking for the
occurrence or the absence of specific subsequences in the strings, then the
(extended) H systems were shown to be very powerful generative mechanisms,
i.e., characterizations of recursively enumerable languages in terms of
various types of H systems were obtained, for example, see \cite{Paun95} and 
\cite{Freundetal95}.

\smallskip

The idea of computations using test tubes as in \cite{Adleman94} (Leonard
Adleman describes the implementation of a small instance of the travelling
salesman problem) was formalized to \textit{test tube systems} using the
splicing operation in \cite{Csuhajetal96}; again, computational completeness
of this computing model could be proved.

The two subprocesses in splicing, i.e., the cutting by enzymes and the
recombination by ligases, were introduced as independent operations in 
\textit{cutting and recombination systems}; computational completeness of
several variants of systems using these operations of cutting and
recombination instead of splicing was exhibited in \cite{FreundWachtler96};
computational completeness of test tube systems using these operations was
proved in \cite{Freundetal96}; computational completeness of H systems using
cut and paste together with other regulation mechanisms as checking for the
occurrence of specific symbols or subsequences was shown, too. In \cite%
{Freunds96}, computational completeness of test tube systems using splicing
or cutting and recombination with the minimal number of two test tubes was
shown; this result is optimal with respect to the number of test tubes,
because due to Dennis Pixton's results from \cite{Pixton2000}, with one test
tube only regular languages can be generated.

For an overview on many interesting models and variants in DNA computing the
interested reader is referred to the monographs \cite{monograph} and \cite%
{handbookDNA}.

\medskip

About ten years ago, another intriguing paradigm based on biology was
introduced by Gheorghe P\u{a}un -- \textit{membrane systems}, soon called 
\textit{P systems} (see \cite{Paun98}); \textit{multisets} of \textit{objects%
} evolve according to \textit{evolution rules} associated with the membranes
arranged in a hierarchical \textit{membrane structure}. A \textit{computation%
} consists of transitions from one \textit{configuration} to the next one,
usually applying the rules in a maximally parallel manner (i.e., applying a
multiset of rules that cannot be extended anymore); the \textit{result} of a
halting computation is given by the objects present in the final
configuration in a specified \textit{output membrane} or by the objects
which leave the external membrane of the system (the \textit{skin} membrane)
during a computation. In tissue(-like) P systems (e.g., see \cite{tiss}) the
membranes are arranged in an arbitrary graph structure instead of a tree
structure as in the original model of P systems. A great variety of variants
has been investigated during the last decade, with the objects being atomic
elements or strings, the rules evolving these objects and/or moving them
through membranes (in P systems) or from one cell to another one (in tissue
P systems). Many models have turned out to be computationally complete, even
with a quite small number of membranes or cells, respectively, and with
quite restricted variants of rules. The interested reader is referred to the
monograph \cite{mcbook} for an introduction to the wide field of (tissue) P
systems and to the P systems web page \cite{Ppage} for the actual state of
the art in P systems.

\medskip

Whereas in P systems and tissue P systems the objects are placed inside the
membranes, in the variant of membrane systems introduced by Luca Cardelli
(see \cite{Brane}), the objects are placed on the membranes. The
computations in these models also called \textit{brane calculus} are based
on specific ways to divide and fuse membranes and to redistribute the
objects on the membranes (e.g., see \cite{Busi}, \cite{NadiaVienna}, \cite%
{projective}), the rules usually being applied in a sequential way in
contrast to the (maximally) parallel way of applying rules in P systems.
Various attempts have already been made to combine different models from the
area of P systems and of brane calculi (e.g., see \cite{CardelliPaun}, \cite%
{credis}). Following this research line by investigating tissue P systems
with the brane operations mate and drip, in \cite{FreundOswaldMecBic}
computational completeness results were obtained both for symbol objects as
well as for string objects. As we shall see later in this paper, the
notations and results given there allow for drawing a close connection to
specific models as investigated in the area of DNA computing and described
above.

\medskip

The rest of the paper is organized as follows: After some preliminary
definitions, we present our definitions for the operations drip and mate and
then show the relation of these operations from the area of (mem)brane
computing with the operations cut and paste used in the area of DNA
computing. In the fourth and in the fifth section, we prove the
computational completeness of test tube systems and of tissue-like P systems
using drip and mate rules working on sets of multisets. A short summary of
results concludes the paper.

\section{Preliminary Definitions}

For the basic elements of formal language theory needed in the following, we
refer to any monograph in this area, in particular to \cite{handbook}. We
just list a few notions and notations: $\mathbb{N}$ denotes the set of
non-negative integers (natural numbers), $\mathbb{N}^{k}$ the set of all $k$%
-vectors of natural numbers. By $\mathbb{N}^{k}RE$ we denote the set of all
recursively enumerable sets of $k$-vectors of natural numbers.

$V^{\ast }$ is the free monoid generated by the alphabet $V$ under the
operation of concatenation; its unit element is the empty string, denoted by 
$\lambda $. The length of a string $x\in V^{\ast }$ is denoted by $|x|$; by $%
RE$ ($RE\left( k\right) $) we denote the family of recursively enumerable
languages (over a $k$-letter alphabet). For any family of string languages $%
F $, $PsF$ denotes the family of Parikh sets of languages from $F$ and $NF$
the family of Parikh sets of languages from $F$ over a one-letter alphabet.
In the following, we will not distinguish between $NRE$, which coincides
with $PsRE\left( 1\right) $, and $RE\left( 1\right) $.

\smallskip

Let $\left\{ a_{1},...,a_{n}\right\} $ be an arbitrary alphabet; the number
of occurrences of a symbol $a_{i}$ in $x$ is denoted by $\left\vert
x\right\vert _{a_{i}};$ the \textit{Parikh vector} associated with $x$ with
respect to $a_{1},...,a_{n}$ is $\left( \left\vert x\right\vert
_{a_{1}},...,\left\vert x\right\vert _{a_{n}}\right) .$ The \textit{Parikh
image} of a language $L$ over $\left\{ a_{1},...,a_{n}\right\} $ is the set
of all Parikh vectors of strings in $L$. For a family of languages $FL,$ the
family of Parikh images of languages in $FL$ is denoted by $PsFL$. A
(finite) multiset $\left\langle m_{1},a_{1}\right\rangle ...\left\langle
m_{n},a_{n}\right\rangle $ with $m_{i}\in \mathbb{N}\mathbf{,}$ $1\leq i\leq
n,$ is represented as any string $x$ the Parikh vector of which with respect
to $a_{1},...,a_{n}$ is $\left( m_{1},...,m_{n}\right) .$

In the following we will not distinguish between a vector $\left(
m_{1},...,m_{n}\right) ,$ its representation by a multiset $\left\langle
m_{1},a_{1}\right\rangle ...\left\langle m_{n},a_{n}\right\rangle $ or its
representation by a string $x$ with Parikh vector $\left( \left\vert
x\right\vert _{a_{1}},...,\left\vert x\right\vert _{a_{n}}\right) =\left(
m_{1},...,m_{n}\right) .$ In that sense, $PsRE\left( k\right) =\mathbb{N}%
^{k}RE.$

\medskip

A \textit{deterministic register machine} is a construct $%
M=(n,B,l_{0},l_{h},I)$, where $n$ is the number of registers, $B$ is a set
of instruction labels, $l_{0}$ is the start label, $l_{h}$ is the halt label
(assigned to \texttt{HALT} only), and $I$ is a set of instructions of the
following forms:

\begin{itemize}
\item $l_{1}:(\mathtt{ADD}(r),l_{2})$ \qquad add 1 to register $r$, and then
go to the instruction labeled by $l_{2}$;

\item $l_{1}:(\mathtt{SUB}(r),l_{2},l_{3})$ \qquad if register $r$ is
non-empty (non-zero), then subtract 1 from it and go to the instruction
labeled by $l_{2}$, otherwise go to the instruction labeled by $l_{3}$;

\item $l_h: \mathtt{HALT}$ \qquad the halt instruction.
\end{itemize}

A deterministic register machine $M$ accepts a set of (vectors of) natural
numbers in the following way: start with the instruction labeled by $l_{0}$,
with the first registers containing the input, as well as all other
registers being empty, and proceed to apply instructions as indicated by the
labels and by the contents of the registers. If we reach the $\mathtt{HALT}$
instruction, then the input number (vector) is accepted. It is known (e.g.,
see \cite{Minsky67}) that in this way we can accept all recursively
enumerable sets of (vectors of) natural numbers. In fact, for accepting any $%
L\in PsRE\left( k\right) $ we need at most $k+2$ registers.

\section{The Operations Mate and Drip}

The reader is supposed to be familiar with basic elements of membrane
computing, (e.g., see the monograph \cite{mcbook} and the P systems web page 
\cite{Ppage}), as well as of brane calculi (see, e.g., \cite{CardelliPaun}).

\smallskip

The operations we are dealing with in this paper are inspired by the ideas
from both areas of P systems and of brane calculi: we consider cells with
the objects being placed on the membranes of the cells (for example, as
already considered in \cite{protmem} and \cite{credis}) -- we will call them 
\textit{vesicles} in the following -- and the operations mate and drip which
are taken from the area of brane calculi and very closely related to the
model of (mem)brane systems already considered in various papers (e.g., see 
\cite{CardelliPaun}, \cite{credis}, \cite{(Mem)brane}), where multisets or
strings (in the biological interpretation we may speak of proteins) are
placed on the membranes. In order to visualize a vesicle with the multiset
of objects $w$ assigned to its membrane we will use the notation ${{[_{{}}}%
_{{}}}_{{}}\ {{]_{{}}}_{{}}}_{w}$ similar to the notation used in the model
of (mem)brane systems.

The two operations drip and mate we shall use in this paper are defined as
follows:

\[
\begin{array}{rl}
drip: & \left( u|c|v;y,z\right) \\ 
mate: & \left( u|a,b|v;x\right)%
\end{array}%
\]

These formal notations describe how to split one cell into two cells (drip)
and how to fuse two cells into one (mate).

Following the notations of \cite{(Mem)brane} used in the model of (mem)brane
systems these operations are interpreted for the concept of vesicles used in
this paper as follows:

The drip operation $\left( u|c|v;y,z\right) $ splits a vesicle (membrane,
cell) ${{[_{{}}}_{{}}}_{{}}\ {{]_{{}}}_{{}}}_{sucvw}$ into the two vesicles $%
{{[_{{}}}_{{}}}_{{}}\ {{]_{{}}}_{{}}}_{suy}$ and ${{[_{{}}}_{{}}}_{{}}\ {{%
]_{{}}}_{{}}}_{zvw}$; $\left( u|a,b|v;x\right) $ fuses a vesicle carrying
the multiset $sua$ and the vesicle carrying the multiset $bvw$ into one
vesicle which then has the multiset $suxvw$, i.e., $ab$ is replaced by $x$
and the remaining multisets are taken as they are. In fact, this means that
from the two vesicles ${{[_{{}}}_{{}}}_{{}}\ {{]_{{}}}_{{}}}_{sua}$ and ${{%
[_{{}}}_{{}}}_{{}}\ {{]_{{}}}_{{}}}_{bvw}$ we get the vesicle ${{[_{{}}}_{{}}%
}_{{}}\ {{]_{{}}}_{{}}}_{suxvw}.$

\smallskip

When dealing with strings, the formal notation is exactly the same as given
above for the case of multisets of objects with the only difference that $%
suy $, $zvw,$ and $sucvw$ have to be interpreted as strings in exactly the
sequence they are written which means that in the case of the drip
operation, we start from a string $sucvw$ which then is split at the site $c$
yielding the two new strings $suy$ and $zvw,$ hence, $s$ and $w$ are not
arbitrary anymore.

\smallskip

In the general case, $a,b,c,s,u,v,w,x,y,z$ can be arbitrary strings over an
alphabet $V$ (no matter whether these are interpreted as multisets of
objects or directly as strings). Computational completeness for tissue P
systems and (mem)brane systems with mate and drip operations working on
strings using a minimal number of membranes was shown in \cite%
{FreundOswaldMecBicentcs} and \cite{FreundOswald06}.

\smallskip

In contrast to this general case which we shall use in this paper, several
restrictions were imposed in \cite{(Mem)brane}:

\begin{enumerate}
\item $a,b,c\in V;$

\item $b=\lambda ,z=\lambda ;$

\item $v\neq \lambda ,ux\neq \lambda .$
\end{enumerate}

As a special variant of the drip rule dealing with a multiset on the skin
membrane of a vesicle we also consider the one-sided drip rule where the
whole rest of the multiset on the membrane of the vesicle to be divided is
put to the first target vesicle, i.e., 
\[
\begin{array}{rl}
drip1: & \left( u|c|v;y,z\right)%
\end{array}%
\]%
which in this case means that from a vesicle ${{[_{{}}}_{{}}}_{{}}\ {{]_{{}}}%
_{{}}}_{sucv}$ we get the two vesicles ${{[_{{}}}_{{}}}_{{}}\ {{]_{{}}}_{{}}}%
_{suy} $ and ${{[_{{}}}_{{}}}_{{}}\ {{]_{{}}}_{{}}}_{vz}$.

\smallskip

In contrast to \cite{(Mem)brane}, where the weight of a drip rule $\left(
u|c|v;y,z\right) $ is defined as the length of the multiset $ucv$ and the
weight of a mate rule $\left( u|a,b|v;x\right) $ as the length of the
multiset $uxv$, we here -- as already considered, for example, in \cite%
{FreundOswaldMecBicentcs} -- define $\left\vert ucvyz\right\vert $ to be the
weight of the drip rule $\left( u|c|v;y,z\right) $ and $|uabvx|$ to be the
weight of a mate rule $\left( u|a,b|v;x\right) $. When using drip rules,
one-sided drip rules, and mate rules of weight at most $k$ we shall write $%
drip_{k}$, $drip1_{k}$, and $mate_{k}$, respectively, as parameters in the
systems (test tube systems and tissue-like P systems) defined in the
succeeding sections.

\subsection{Relating DNA Computing and Membrane Computing}

As already exhibited in \cite{Freund08}, we may observe a coincidence with
operations well known from the area of DNA computing when looking carefully
into the definitions of the operations mate and drip and the results of
applying them to strings: in \cite{FreundWachtler96}, the operations \textit{%
cutting and recombination} of strings, operations which are closely related
to the splicing operation, were considered; as we shall exhibit in the
following lines, cutting respectively its more general variant \textit{cut}
is similar to the operation drip and recombination respectively its more
general variant \textit{paste} is similar to the operation mate.

\smallskip

The \textit{cutting} operation means cutting a string into two pieces, with
adding strings on the cutting sites of the cut pieces; the \textit{%
recombination} operation means fusing two strings thereby eliminating
substrings at the fusion sites of both strings. The substrings added at the
cutting sites and those eliminated at the fusion sites can be interpreted,
for example, as electrical charges of molecules.

More general variants are the \textit{cut} and \textit{paste} operations
formally to be written as follows: 
\[
\begin{array}{rl}
cut: & \left( u|c|v;y,z\right)  \\ 
& \mbox{cut one string into two strings} \\ 
paste: & \left( u|a,b|v;x\right)  \\ 
& \mbox{recombine two strings into one}%
\end{array}%
\]%
The cut operation $\left( u|c|v;y,z\right) $ means splitting one string into
two strings: a string $sucvw$ is split into the two strings $suy$ and $zvw$,
i.e., $c$ is eliminated and replaced by $y$ at the end of the first
substring and by $z$ at the beginning of the second substring; formally this
can be written as $sucvw\Longrightarrow \left( suy,zvw\right) $. The paste
operation $\left( u|a,b|v;x\right) $ means fusing two strings to one string:
a string $sua$ and a string $bvw$ are fused to the single string $suxvw$,
i.e., $ab$ is replaced by $x$ and the remaining substrings are taken as they
are; formally this can be written as $\left( sua,bvw\right) \Longrightarrow
sucvw$. In \textit{cutting and recombination} systems, we have the
restrictions $x=\lambda $ and $c=\lambda $.

\medskip

Looking carefully into these notations of the operations cut and paste as
well as drip and mate and the effect of applying them to strings or
multisets, we realize that we have got \textit{identical notations:} 
\[
\begin{array}{rl}
mate/paste: & \left( u|a,b|v;x\right) \\ 
drip/cut: & \left( u|c|v;y,z\right)%
\end{array}%
\]%
With respect to the interpretation in tissue P systems with mate and drip
operations, a string assigned to a cell corresponds with this string itself
in the interpretation of DNA computing. Hence, we observe that the\textit{\
mate and drip operations }and the \textit{cut and paste operations }are
closely related. In that way, results established and questions/problems
raised for systems using the mate and drip operations may also be
established/raised for the corresponding systems using the cut and paste
(cutting and recombination) operations and vice versa.

As a specific example of relating the two areas of DNA computing and
membrane computing, we take over the idea of working with sets from DNA
computing instead of working with multisets as usually done in the area of
membrane computing to the model of \textit{tissue-like P systems with mate
and drip rules.} On the other hand, we will use the drip and mate rules in
test tube systems working on multisets of elementary objects placed on
membranes.

\section{Test Tube Systems with Drip and Mate Rules}

In this section, we prove our first main result establishing the
computational completeness of variants of test tube systems with mate and
drip rules working on sets of multisets, i.e., as objects in the test tubes
we consider sets of vesicles carrying multisets of elementary objects
(symbols) on their skin membrane, and as operations acting in the test tubes
we take the operations drip and mate processing these vesicles.

\medskip

We use the following general definition for test tube systems as in \cite%
{Freunds96}, where the contents of the tubes is redistributed to selected
test tubes according to specific filters: \smallskip

A \textit{test tube system} (a \textit{TTS} for short) $\sigma $ is a
construct 
\[
\left( O,O_{T},n,A,\rho ,D,E\right) 
\]%
where

\begin{enumerate}
\item $O$ is a set of \textit{objects;}

\item $O_{T}$ is a set of \textit{terminal objects, }$O_{T}\subseteq O$;

\item $n$, $n\geq 1$, is the number of test tubes in $\sigma $;

\item $A=\left( A_{1},...,A_{n}\right) $ is a sequence of sets of \textit{%
axioms,} where $A_{i}\subseteq O,$ $1\leq i\leq n;$

\item $\rho $ is a sequence $\left( \rho _{1},...,\rho _{n}\right) $ of sets
of \textit{test tube operations,} where $\rho _{i}$ contains specific
operations for the test tube $T_{i}$, $1\leq i\leq n$;

\item $D$ is a (finite) set of \textit{prescribed output/input relations}
between the test tubes in $\sigma $ of the form $\left( i,F,j\right) ,$
where $1\leq i\leq n$, $1\leq j\leq n$, $i\neq j$, and $F$ is a (recursive)
subset of $O$; $F$ is called a filter between the test tubes $T_{i}$ and $%
T_{j}$;

\item $E\subseteq \left\{ i\mid 1\leq i\leq n\right\} $ specifies the set of 
\textit{output tubes}.
\end{enumerate}

In the interpretation used in \cite{Freunds96}, the computations in the
system $\sigma $ run as follows: At the beginning of each computation step
the axioms are distributed over the $n$ test tubes according to $A$, hence,
test tube $T_{i}$ starts its first computation step with $A_{i}$. Now let $%
L_{i}$ be the contents of test tube $T_{i}$ at the beginning of a
computation step. Then in each test tube the rules of $\rho _{i}$ operate on 
$L_{i}$, i.e., we obtain $\rho _{i}^{\ast }\left( L_{i}\right) $, where $%
\rho _{i}^{\ast }\left( L_{i}\right) =\cup _{n=0}^{\infty }\rho _{i}^{\left(
n\right) }\left( L_{i}\right) $ with $\rho _{i}^{\left( n\right) }\left(
L_{i}\right) $ being defining inductively as follows: $\rho _{i}^{\left(
0\right) }\left( L_{i}\right) =L_{i}$ and $\rho _{i}^{\left( n+1\right)
}\left( L_{i}\right) =\rho _{i}^{\left( n\right) }\left( L_{i}\right) \cup
\rho _{i}\left( \rho _{i}^{\left( n\right) }\left( L_{i}\right) \right) $
for $n\geq 0$; for any set $L$, $\rho _{i}\left( L\right) $ is the set of
all objects obtained by applying rules from $\rho _{i}$ to objects from $L$.
The next substep is the redistribution of the $\rho _{i}^{\ast }\left(
L_{i}\right) $ over all test tubes according to the corresponding
output/input relations from $D$, i.e., if $\left( i,F,j\right) \in D$, then
the test tube $T_{j}$ from $\rho _{i}^{\ast }\left( L_{i}\right) $ gets $%
\rho _{i}^{\ast }\left( L_{i}\right) $ whereas the rest of $\rho _{i}^{\ast
}\left( L_{i}\right) $ that cannot be distributed to other test tubes
remains in $T_{i}$. The final result of the computations in $\sigma $
consists of all terminal objects from $O_{T}$ that can be extracted from an 
\textit{output tube} $f$ from $E,$ i.e., we take $\rho _{f}^{\ast }\left(
L_{f}\right) \cap O_{T}$.

In this paper, we allow a more relaxed view of processing the operations in
the test tubes and the succeeding redistribution of the objects therein,
i.e., we assume that at any moment objects fulfilling the specific
constraints given by a filter $\left( i,F,j\right) \in D$ may pass from test
tube $T_{i}$ to test tube $T_{j}$, with some copies remaining in $T_{i}$. In
the limit, the same results can be obtained in that way as in the strict
interpretation as described before, yet our more relaxed interpretation
allows for a much easier description of development of objects as will be
seen in the following.

The multisets only consisting of terminal objects found on vesicles in an
output tube form the set of results generated by a test tube system, and the
family of all such sets of multisets over a terminal alphabet with
cardinality $k$ generated by test tube systems using at most $m$ test tubes,
axioms of weight at most $l$, drip rules of weight at most $q$, and mate
rules of weight at most $p$ is denoted by 
\[
TTS_{m}\left( axiom_{l},drip_{q},mate_{p}\right) \left( k\right) =PsRE\left(
k\right) . 
\]

\bigskip

\noindent \textit{Theorem 1.} $TTS_{m}\left( axiom_{l},mate_{p}\right)
\left( k\right) =PsRE\left( k\right) $ for all $m\geq 3$, $l\geq 3$, $p\geq
5 $, $k\geq 1$.

\bigskip

\noindent \textit{Proof.} Let $M=(n,B,p_{0},p_{h},I)$ be a register machine
with $n$ registers accepting $L\in PsRE\left( k\right) $; moreover, let $%
B_{ADD}$ and $B_{SUB}$ denote the sets of labels of the ADD- and
SUB-instructions in $I$, respectively, i.e.,%
\[
\begin{array}[t]{rrl}
B_{ADD} & = & \left\{ l_{1}\mid l_{1}:\left( \mathtt{ADD}\left( r\right)
,l_{2}\right) \in I\right\} , \\ 
B_{SUB} & = & \left\{ l_{1}\mid l_{1}:\left( \mathtt{SUB}\left( r\right)
,l_{2},l_{3}\right) \in I\right\} .%
\end{array}%
\]%
Then we construct a TTS $\sigma $ 
\[
\left( O,O_{T},3,A,\rho ,D,\left\{ 3\right\} \right) 
\]%
with three test tubes and mate rules of weight five generating $L$, with the
contents of register $i$ represented as the number of symbols $b_{i}$ as
follows:

The objects in $O$ are vesicles of the form ${{[_{{}}}_{{}}}_{{}}\ {{]_{{}}}%
_{{}}}_{w}$ with $w$ being a multiset over an alphabet $V$ to be specified
below; yet we may simply represent such an object by any string representing 
$w$; hence, we can also write $O=V^{\ast }$ where 
\[
\begin{array}{ccl}
V & = & B\cup \left\{ X,Y,Z,F\right\} \cup \left\{ a_{i}\mid 1\leq i\leq
k\right\} \cup \left\{ b_{i}\mid 1\leq i\leq n\right\} \\ 
& \cup & \left\{ A_{l}\mid l\in B_{ADD}\right\} \cup \left\{
A_{l},A_{l}^{\prime },A_{l}^{\prime \prime }\mid l\in B_{SUB}\right\} .%
\end{array}%
\]%
In the same sense, we will write $O_{T}=V_{T}^{\ast }$ with $V_{T}=\left\{
a_{i}\mid 1\leq i\leq k\right\} $.

In the first test tube $T_{1}$, we initialize the simulation of a
computation in the register machine $M$ with obtaining (vesicles carrying)
multisets of the form $%
Xa_{1}^{n_{1}}...a_{k}^{n_{k}}b_{1}^{n_{1}}...b_{k}^{n_{k}}$ using the
axioms 
\[
\left\{ X,Zl_{0}\right\} \cup \left\{ a_{i}b_{i}Y\mid 1\leq i\leq k\right\} 
\]%
and the mate rules\ $(X\mid ,Y\mid ;)$ and $(X\mid ,Z\mid l_{0};)$; with
applying the second rule, we start the simulation of a computation in the
register machine $M$.

Moreover, $l_{1}:(\mathtt{ADD}(r),l_{2})\in I$ is simulated by the axiom $%
A_{l_{1}}l_{2}b_{r}$ and the mate rule $\left( X\mid l_{1},A_{l_{1}}\mid
l_{2}b_{r};\right) $.

For $l_{1}:(\mathtt{SUB}(r),l_{2},l_{3})\in I$, the subtract case is
simulated using the axiom\ $A_{l_{1}}l_{2}$ and the mate rule$\ (X\mid
l_{1}b_{r},\ A_{l_{1}}\mid l_{2};)$. The case when we guess the contents of
register $r$ to be zero is started with using the axiom\ $A_{l_{1}}^{\prime }
$ together with the mate rule$\ (X\mid l_{1},\mid A_{l_{1}}^{\prime };)$.
The computation is then continued in test tube $T_{2}$ where the rule $%
(X\mid A_{l_{1}}^{\prime },A_{l_{1}}^{\prime \prime }\mid l_{3};)$ with the
axiom\ $A_{l_{1}}^{\prime \prime }l_{3}$ allows for sending back the
multiset in case that the guess has been correct. Appearance checking
(testing that no symbol $b_{r}$ is present) in the zero case for $l_{1}:(%
\mathtt{SUB}(r),l_{2},l_{3})\in I$ is accomplished by the corresponding
filter in 
\[
\left( 1,\cup _{1\leq r\leq n}(V_{T}\cup \left\{ X\right\} \cup \left\{
b_{i}\mid 1\leq i\leq n,i\neq r\right\} \cup \left\{ A_{l_{1}}^{\prime }\mid
l_{1}:(\mathtt{SUB}(r),l_{2},l_{3})\in I\right\} )^{\ast },2\right) 
\]%
from test tube $T_{1}$ to test tube $T_{2}$ and$\ $%
\[
(2,\left( V\ -\left\{ A_{l}^{\prime },A_{l}^{\prime \prime }\mid l\in
B_{SUB}\right\} \right) ^{\ast },1)
\]%
from test tube $T_{2}$ back to test tube $T_{1}$.

The terminal results are collected in test tube $T_{3}$ by eliminating the
symbol $X$ which is present in every multiset representing a configuration
of a computation in $M$ as soon as the final label $l_{h}$ has appeared with
using the mate rule\ $(\mid l_{h}X,F\mid ;)$ with the axiom\ $F$ in test
tube $T_{1}$ and then letting these terminal multisets get through the
filter $\left( 1,\left\{ a_{i}\mid 1\leq i\leq k\right\} ^{\ast },3\right) $
from test tube $T_{1}$ to test tube $T_{3}$.

The sets of axioms, rules, and prescribed output/input relations (filters) $A
$, $\rho $, and $D$, respectively, can easily be collected from the
descriptions given above:%
\[
\begin{array}[t]{rll}
A & = & \left( A_{1},A_{2},\emptyset \right) , \\ 
A_{1} & = & \left\{ X,Zl_{0},F\right\} \cup \left\{ a_{i}b_{i}Y\mid 1\leq
i\leq k\right\}  \\ 
& \cup  & \left\{ A_{l_{1}}l_{2}b_{r}\mid l_{1}:(\mathtt{ADD}(r),l_{2})\in
I\right\}  \\ 
& \cup  & \left\{ A_{l_{1}}l_{2},A_{l_{1}}^{\prime }\mid l_{1}:(\mathtt{SUB}%
(r),l_{2},l_{3})\in I\right\} , \\ 
A_{2} & = & \left\{ A_{l_{1}}^{\prime \prime }l_{3}\mid l_{1}:(\mathtt{SUB}%
(r),l_{2},l_{3})\in I\right\} , \\ 
\rho  & = & \left( \rho _{1},\rho _{2},\emptyset \right) , \\ 
\rho _{1} & = & \left\{ (X\mid ,Y\mid ;),(X\mid ,Z\mid l_{0};),(\mid
l_{h}X,F\mid ;)\right\}  \\ 
& \cup  & \left\{ \left( X\mid l_{1},A_{l_{1}}\mid l_{2}b_{r};\right) \mid
l_{1}:(\mathtt{ADD}(r),l_{2})\in I\right\}  \\ 
& \cup  & \left\{ (X\mid l_{1}b_{r},\ A_{l_{1}}\mid l_{2};),(X\mid
l_{1},\mid A_{l_{1}}^{\prime };)\mid l_{1}:(\mathtt{SUB}(r),l_{2},l_{3})\in
I\right\} , \\ 
\rho _{2} & = & \left\{ (X\mid A_{l_{1}}^{\prime },A_{l_{1}}^{\prime \prime
}\mid l_{3};)\mid l_{1}:(\mathtt{SUB}(r),l_{2},l_{3})\in I\right\} , \\ 
D & = & \{(1,\cup _{1\leq r\leq n}(V_{T}\cup \left\{ X\right\} \cup \left\{
b_{i}\mid 1\leq i\leq n,i\neq r\right\} \cup \{A_{l_{1}}^{\prime }\mid
l_{1}:(\mathtt{SUB}(r),l_{2},l_{3})\in I\})^{\ast },2), \\ 
&  & \hspace*{0.2cm}(2,\left( V\ -\left\{ A_{l}^{\prime },A_{l}^{\prime
\prime }\mid l\in B_{SUB}\right\} \right) ^{\ast },1),(1,\left\{ a_{i}\mid
1\leq i\leq k\right\} ^{\ast },3)\}.%
\end{array}%
\]

As desired, we use only three test tubes, axioms of weight at most three,
and mate rules of weight at most five; moreover, the filters in the
prescribed output/input relations of the TTS $\sigma $ are of the very
special and simple form $\left( i,W^{\ast },j\right) $ with $W\subset V$ or
finite unions of such filters. These observations complete the proof.\hfill $%
\Box $

\bigskip

As an alternative to having all the axioms in the test tubes as indicated in
the proof constructed above, we may use the single axiom $g$ and\ the drip
rule

\[
(\mid g\mid ;A,) 
\]%
for each axiom $A$. Hence, we immediately obtain the following result:

\bigskip

\noindent \textit{Corollary 2.} $TTS_{m}\left(
axiom_{l},drip_{q},mate_{p}\right) \left( k\right) =PsRE\left( k\right) $
for all $m\geq 3$, $l\geq 1$, $p\geq 5$, $q\geq 4$, $k\geq 1$.

\bigskip

\noindent \textit{Proof.} All required axioms can be computed from the
single axiom $g$ by using\ the drip rule $(\mid g\mid ;A,)$ -- as well as by
using $(\mid g\mid ;g,)$ for $g$ itself -- in each of the two test tubes $%
T_{1}$ and $T_{2}$. As a small technical detail we mention that the
computations in these new test tube systems need an additional step at the
beginning to initialize the two test tubes $T_{1}$ and $T_{2}$ with the
corrsponding set of axioms. \hfill $\Box $

\bigskip

Another interesting variant is the use of one-sided drip rules instead of
mate rules: looking carefully into the proof of Theorem~1 and the mate rules
used there we realize that the second vesicle always carries an axiom. In
general, if $bv$ is the whole second vesicle, then the mate rule

\[
\left( u\mid a,b\mid v;x\right) 
\]%
can be simulated by the one-sided drip rule 
\[
(u\mid a\mid ;vx,), 
\]%
i.e., we put everything to the first vesicle and thus in fact obtain only
one result by applying this rule.

\bigskip

\noindent \textit{Corollary 3.} $TTS_{m}\left( axiom_{l},drip1_{q}\right)
\left( k\right) =PsRE\left( k\right) $ for all $m\geq 3$, $l\geq 1$, $q\geq
4 $, $k\geq 1$.

\bigskip

\noindent \textit{Proof.} According to the proof of Corollary~2, we can get
every axiom by a one-sided drip rule. Moreover, as explained above, every
mate rule $\left( u\mid a,b\mid v;x\right) $ used in the proof of Theorem~1
can be replaced by the one-sided drip rule $(u\mid a\mid ;vx,)$.\hfill $\Box 
$

\section{Tissue-like P Systems with Mate and Drip Rules}

In this section, we prove our main result establishing the computational
completeness of variants of tissue-like P systems with mate and drip rules
working on sets of multisets.

\medskip

A \textit{tissue-like P systems with mate and drip rules} (\textit{tP system}
for short) $\Pi $ is a construct 
\[
\left( V,V_{T},n,A,R,i_{0}\right) 
\]%
where

\begin{enumerate}
\item $V$ is a finite set of \textit{symbols;}

\item $V_{T}$ is a set of \textit{terminal symbols, }$V_{T}\subseteq V$;

\item $n$, $n\geq 1$, is the number of cells in $\Pi $;

\item $A=\left( A_{1},...,A_{n}\right) $ is a sequence of sets of \textit{%
axioms,} where $A_{i}\subseteq V^{\ast }$, $1\leq i\leq n$, describing the
initial contents of the cells;

\item $R$ is a set of \textit{rules} of the form 
\[
T_{i}:r\rightarrow T_{j}
\]%
with $i,j\in \left\{ l\mid 1\leq l\leq n\right\} $, $i\neq j$, and $r$ being
a drip or mate rule over $V$;

\item $i_{0}\in \left\{ l\mid 1\leq l\leq n\right\} $ specifies the \textit{%
output cell}.
\end{enumerate}

\bigskip

A computation in $\Pi $ starts with the initial configuration described by $%
A $; a computation step then consists of applying the rules $%
T_{i}:r\rightarrow T_{j}$ in the $i$-th cell -- the application of a rule $%
T_{i}:r\rightarrow T_{j}$ means applying $r$ to objects in (the source) cell 
$T_{i}$ and sending the resulting vesicle(s) to (the target) cell $T_{j}$ --
in a maximal way in that sense that every vesicle that can undergo the
application of a rule will be affected by a suitable rule, yet as we are
dealing with sets of vesicles, this also means that any vesicle or any pair
of vesicles has to be used with every possible rule by which it can be
affected.

The multisets only consisting of terminal objects found on vesicles in the
output cell $i_{0}$ form the set of results generated by $\Pi $, and the
family of all such sets of multisets over a terminal alphabet with
cardinality $k$ generated by tissue-like P systems using at most $m$ cells,
axioms of weight at most $l$, drip rules of weight at most $q$, and mate
rules of weight at most $p$ is denoted by 
\[
tP_{m}\left( axiom_{l},drip_{q},mate_{p}\right) \left( k\right) =PsRE\left(
k\right) . 
\]

\bigskip

\noindent \textit{Theorem 4.} $tP_{m}\left(
axiom_{l},drip_{q},mate_{p}\right) \left( k\right) =PsRE\left( k\right) $
for all $m\geq 5$, $l\geq 3$, $p\geq 5$, $q\geq 5$, $k\geq 1$.

\bigskip

\noindent \textit{Proof.} Let $M=(n,B,p_{0},p_{h},I)$ be a register machine
with $n$ registers accepting $L\in PsRE\left( k\right) $; then we construct
a tissue-like P system $\Pi $ 
\[
\left( V,V_{T},5,A,R,5\right) 
\]%
generating $L$. We start with the following initial vesicles in the five
cells:

\[
\begin{array}{ccl}
A_{1} & = & \{B_{s}\mid s\in \left\{ X,Zl_{0},F\right\} \cup \left\{
a_{i}b_{i}Y\mid 1\leq i\leq k\right\}  \\ 
&  & \hspace*{1.45cm}\cup \left\{ A_{l_{1}}l_{2}b_{r}\mid l_{1}:\left( 
\mathtt{ADD}\left( r\right) ,l_{2}\right) \in I\right\}  \\ 
&  & \hspace*{1.45cm}\cup \left\{ A_{l_{1}}l_{2},A_{l_{1}}^{\prime }\mid
l_{1}:\left( \mathtt{SUB}\left( r\right) ,l_{2},l_{3}\right) \in I\right\}
\}, \\ 
A_{2} & = & \emptyset , \\ 
A_{3} & = & \left\{ E_{r}l_{3},F_{r}D_{r}\mid l_{1}:\left( \mathtt{SUB}%
\left( r\right) ,l_{2},l_{3}\right) \in I\right\} , \\ 
A_{4} & = & \left\{ A_{r}\mid l_{1}:\left( \mathtt{SUB}\left( r\right)
,l_{2},l_{3}\right) \in I\right\} , \\ 
A_{5} & = & \emptyset .%
\end{array}%
\]

In general, for generating a multiset $s$ in the first cell $T_{1}$ we use
the following rules in $T_{1}$ and $T_{2}$:

$T_{1}:(\mid B_{s}\mid ;R,B_{s}^{\prime }B_{s}s)\rightarrow T_{2}$,

$T_{2}:(\mid R,B_{s}^{\prime }B_{s}\mid ;)\rightarrow T_{1}$ generates $s$
in $T_{1}$,

$T_{2}:(\mid R,B_{s}^{\prime }s\mid ;)\rightarrow T_{1}$ regains $B_{s}$ in $%
T_{1}$.

Moreover, for sending back from $T_{2}$ to $T_{1}$ a multiset containing the
specific symbol $X$ indicating a multiset on a vesicle representing a
configuration of a computation in the register machine $M$, we use the
special symbol $R$ with the rule

$T_{2}:(X\mid ,R\mid ;)\rightarrow T_{1}$.

For the initialization as already explained in the proof of Theorem~1, we
take $s=X$, $s=a_{i}b_{i}Y$ for $1\leq i\leq k$, and $s=Zl_{0}\ $as well as
the rules\ 

$T_{1}:(X\mid ,Y\mid ;)\rightarrow T_{2}$ and

$T_{1}:(X\mid ,Z\mid l_{0};)\rightarrow T_{2}$;

\noindent with applying the second rule, we start the simulation of a
computation in the register machine $M$.

For simulating an ADD-instruction $l_{1}:(\mathtt{ADD}(r),l_{2})\in I$ we
take $s=A_{l_{1}}l_{2}b_{r}$ and the rule

$T_{1}:(X\mid l_{1},A_{l_{1}}\mid l_{2}b_{r};)\rightarrow T_{2}$.

For simulating a SUB-instruction $l_{1}:(\mathtt{SUB}(r),l_{2},l_{3})\in I$
in the case that subtraction\ is possible we take $s=A_{l_{1}}l_{2}$ and the
rule

$T_{1}:(X\mid l_{1}b_{r},A_{l_{1}}\mid l_{2};)\rightarrow T_{2}$.

In all the cases described so far, the main work is done by a rule of the
form $T_{1}:r\rightarrow T_{2}$ using a rule in $T_{1}$ with the result
being sent to cell $T_{2}$, where with the application of the rule

$T_{2}:(X\mid ,R\mid ;)\rightarrow T_{1}$

\noindent we already described before, the result is sent back to cell $%
T_{1} $.

For simulating a SUB-instruction $l_{1}:(\mathtt{SUB}(r),l_{2},l_{3})\in I$
in the case that subtraction\ is not possible we take $s=A_{l_{1}}^{\prime }$
and guess that no $b_{r}$ occurs, but now send the result to cell $T_{3}$:

$T_{1}:(X\mid l_{1},\mid A_{l_{1}}^{\prime };)\rightarrow T_{3}$.

Checking for the occurrence of $b_{r}$ now is accomplished by the following
rules affecting a vesicle containing $X$ within a cycle of $2$; in even
computation steps, the rule $T_{3}:\left( \mid B_{r},X\mid b_{r};\right)
\rightarrow T_{4}$ \textquotedblleft kills\textquotedblright\ vesicles
containing $b_{r}$ by sending them to cell $T_{4}$ thereby also erasing the
symbol $X$ so that it cannot be affected by a rule anymore. If no $b_{r}$
occurs, then one step later the rule $T_{3}:\left( \mid A_{l_{1}}^{\prime
},E_{r}\mid l_{3};\right) \rightarrow T_{2}$ sends the vesicle with the desired label $l_{3}$ back to cell $T_{1}$ via
cell $T_{2}$ (hence, in total the simulation of this case takes four steps).
The symbols $A_{r},B_{r},C_{r}$ and $D_{r},E_{r},F_{r}$, respectively, allow
for having the desired checking symbols $B_{r}$ and $E_{r}$ in $T_{3}$ at
the right moment, i.e., if a vesicle has \textquotedblleft
survived\textquotedblright\ $B_{r}$, then $E_{r}$ will finish the simulation
of the zero-case of the SUB-instruction.

$T_{4}:\left( \mid A_{r}\mid ;B_{r},C_{r}A_{r}\right) \rightarrow T_{3},$

$T_{3}:\left( \mid B_{r},C_{r}\mid A_{r};\right) \rightarrow T_{4},$

$T_{3}:\left( \mid B_{r},X\mid b_{r};\right) \rightarrow T_{4},$

$T_{4}:\left( \mid D_{r}\mid ;E_{r}l_{3},F_{r}D_{r}\right) \rightarrow
T_{3}, $ for $l_{1}:\left( \mathtt{SUB}\left( r\right) ,l_{2},l_{3}\right)
\in I,$

$T_{3}:\left( \mid E_{r}l_{3},F_{r}\mid D_{r};\right) \rightarrow T_{4},$

$T_{3}:\left( \mid A_{l_{1}}^{\prime },E_{r}\mid l_{3};\right) \rightarrow
T_{2},$ for $l_{1}:\left( \mathtt{SUB}\left( r\right) ,l_{2},l_{3}\right)
\in I.$

To obtain the output vesicles in $T_{5}$, we apply the rule

$T_{1}:\left( l_{h}X\mid ,F\mid ;\right) \rightarrow T_{5}$.

In sum, we obtain the following set of rules $R$:
\[
\begin{array}[t]{lll}
R & = & \{T_{1}:(\mid B_{s}\mid ;R,B_{s}^{\prime }B_{s}s)\rightarrow T_{2},
\\ 
&  & \hspace*{0.2cm}T_{1}:(\mid B_{s}\mid ;R,B_{s}^{\prime
}B_{s}s)\rightarrow T_{2}, \\ 
&  & \hspace*{0.2cm}T_{2}:(\mid R,B_{s}^{\prime }s\mid ;)\rightarrow T_{1}
\\ 
&  & \hspace*{0.2cm}\mid s\in \left\{ X,Zl_{0},F\right\} \cup \left\{
a_{i}b_{i}Y\mid 1\leq i\leq k\right\}  \\ 
&  & \hspace*{0.7cm}\cup \left\{ A_{l_{1}}l_{2}b_{r}\mid l_{1}:(\mathtt{ADD}%
(r),l_{2})\in I\right\}  \\ 
&  & \hspace*{0.7cm}\cup \left\{ A_{l_{1}}l_{2},A_{l_{1}}^{\prime }\mid
l_{1}:(\mathtt{SUB}(r),l_{2},l_{3})\in I\right\} \} \\ 
& \cup  & \left\{ T_{1}:(X\mid ,Y\mid ;)\rightarrow T_{2},T_{1}:(X\mid
,Z\mid l_{0};)\rightarrow T_{2}\right\}  \\ 
& \cup  & \left\{ T_{2}:(X\mid ,R\mid ;)\rightarrow T_{1},T_{1}:\left(
l_{h}X\mid ,F\mid ;\right) \rightarrow T_{5}\right\}  \\ 
& \cup  & \{T_{1}:(X\mid l_{1},A_{l_{1}}\mid l_{2}b_{r};)\rightarrow
T_{2}\mid l_{1}:(\mathtt{ADD}(r),l_{2})\in I\} \\ 
& \cup  & \{T_{1}:(X\mid l_{1},\mid A_{l_{1}}^{\prime };)\rightarrow
T_{3},T_{1}:(X\mid l_{1}b_{r},A_{l_{1}}\mid l_{2};)\rightarrow T_{2}, \\ 
&  & \hspace*{0.2cm}T_{3}:\left( \mid B_{r},X\mid b_{r};\right) \rightarrow
T_{4},T_{3}:\left( \mid A_{l_{1}}^{\prime },E_{r}\mid l_{3};\right)
\rightarrow T_{2}, \\ 
&  & \hspace*{0.2cm}T_{4}:\left( \mid A_{r}\mid ;B_{r},C_{r}A_{r}\right)
\rightarrow T_{3},T_{3}:\left( \mid B_{r},C_{r}\mid A_{r};\right)
\rightarrow T_{4}, \\ 
&  & \hspace*{0.2cm}T_{4}:\left( \mid D_{r}\mid
;E_{r}l_{3},F_{r}D_{r}\right) \rightarrow T_{3},T_{3}:\left( \mid
E_{r}l_{3},F_{r}\mid D_{r};\right) \rightarrow T_{4}, \\ 
&  & \hspace*{0.2cm}T_{3}:\left( \mid A_{l_{1}}^{\prime },E_{r}\mid
l_{3};\right) \rightarrow T_{2}\mid l_{1}:(\mathtt{SUB}(r),l_{2},l_{3})\in
I\}%
\end{array}%
\]

We emphasize once more that the simulation of any computation step of the
register machine $M$ takes an even number of steps (i.e., two or four), and
also in the initial phase, i.e., the generation of the axioms and the
initial configurations $Xwl_{0}$ with $w\in \left\{ a_{i}b_{i}\mid 1\leq
i\leq k\right\} ^{\ast }$ in the first cell $T_{1}$ takes an even number of
steps, which guarantees that the zero-check performed by the interplay of
rules in the cells $T_{3}$ and $T_{4}$ works correctly. Finally, we mention
the computation in $\Pi $ never stops and every element of $L$ will appear
as the multiset on a vesicle in the output cell at some moment during the
computation in $\Pi $ and will be sent to cell $T_{5}$ again in each odd
step of the computation after its first appearance in $T_{5}$, as every
computation of the register machine $M$ can be started again after any even
number of computation steps in $\Pi $. These observations complete the
proof. \hfill $\Box $

\section{Conclusion}

As in DNA computing, we have considered sets of objects instead of multisets
as mostly considered in the area of P systems. The operations cut and
recombination well known from DNA computing have their counterparts as the
operations drip and mate considered in (mem)brane computing. We have
investigated the computational power of specific variants of the operations
drip and mate on sets of vesicles with multisets of objects on their outside
membrane acting in test tube systems, where the vesicles pass from one tube
to another one provided they fulfill specific constraints, and in
tissue-like P systems, where the vesicles are passed to specified cells
after having undergone a drip or mate operation. In both setups, we have
proved computational completeness, even with different variants of the drip
and mate operations. As far as the descriptional complexity of the test tube
systems with respect to the number of test tubes and of the tissue-like P
systems with respect to the number of cells and in both cases with respect
to the weight of the mate and drip operations is concerned, improving the
obtained results in these respects remains as a challenging task for future
research.

\bigskip

\noindent \textbf{Acknowledgements:} The first author gratefully
acknowledges many interesting discussions with Gheorghe P\u{a}un and Marion
Oswald on several topics considered in this paper.

\end{document}